\documentclass[aps,prl,amsmath,superscriptaddress,twocolumn,longbibliography]{revtex4-1}

\usepackage{graphicx}
\usepackage{amsmath}
\usepackage{amsfonts}
\usepackage{amssymb}
\usepackage{braket}
\usepackage{bm}
\usepackage{multirow}
\usepackage{color}
\usepackage[normalem]{ulem}
\usepackage{mathrsfs}
\usepackage{mathtools}
\usepackage{float}

\makeatletter

\newcommand{\Rmnum}[1]{\expandafter\@slowromancap\romannumeral #1@}
\makeatother

\usepackage[breaklinks]{hyperref}
\hypersetup{colorlinks=true, linkcolor=blue, citecolor=blue, filecolor=blue, urlcolor=blue}

\AtBeginDocument{%
    \newwrite\bibnotes
    \def\bibnotesext{Notes.bib}
    \immediate\openout\bibnotes=\jobname\bibnotesext
    \immediate\write\bibnotes{@CONTROL{REVTEX41Control}}
    \immediate\write\bibnotes{@CONTROL{%
    apsrev41Control,author="08",editor="1",pages="1",title="0",year="1"}}
     \if@filesw
     \immediate\write\@auxout{\string\citation{apsrev41Control}}%
    \fi
}%

\def\sz{\sigma^{\rm z}}
\def\sx{\sigma^{\rm x}}
\def\mz{\mu^{\rm z}}
\def\mx{\mu^{\rm x}}
\def\tz{\tau^{\rm z}}
\def\tx{\tau^{\rm x}}
\def\da{\downarrow}
\def\ua{\uparrow}

\begin{document}

\title{Constructing Quantum Spin Liquids Using Combinatorial Gauge Symmetry}

\author{Claudio Chamon}
\email{chamon@bu.edu}
\thanks{These authors contributed equally}
\affiliation{Physics Department, Boston University, Boston, MA, 02215, USA}

\author{Dmitry Green}
\email{dmitry.green@aya.yale.edu}
\thanks{These authors contributed equally}
\affiliation{AppliedTQC.com, ResearchPULSE LLC, New York, NY 10065, USA}

\author{Zhi-Cheng Yang}
\email{yangzc@bu.edu}
\affiliation{Physics Department, Boston University, Boston, MA, 02215, USA}

\date{\today}

\begin{abstract}
We introduce the notion of combinatorial gauge symmetry -- a local
transformation that includes single spin rotations plus permutations
of spins (or swaps of their quantum states) -- that preserve the
commutation and anti-commutation relations among the spins. We show
that Hamiltonians with simple two-body interactions contain this
symmetry if the coupling matrix is a Hadamard matrix, with the
combinatorial gauge symmetry being associated to the automorphism of
these matrices with respect to monomial transformations. Armed with
this symmetry, we address the physical problem of how to build quantum
spin liquids with physically accessible interactions. In addition to
its intrinsic physical significance, the problem is also tied to that
of how to build topological qubits.
\end{abstract}

\maketitle


Quantum liquids of spins are systems where no magnetic
symmetry-breaking order should be detectable down to zero
temperature~\cite{Balents2010}, and instead topological order
exits~\cite{Wen1990a}. On the theoretical side, there are a number of
model Hamiltonians where quantum spin liquid states
exist~\cite{tqc_Kitaev,hex_Kitaev}. Gauge symmetries are common in
these models, whether discrete or continuous, intrinsic or
emergent. Many of these gauge models, such as the ${\mathbb Z}_2$
toric code~\cite{tqc_Kitaev} and fracton models such as the
$X$-cube~\cite{PhysRevB.81.184303, PhysRevB.94.235157}, are defined
using multi-spin interactions. Here, we show that {\it exact} local
${\mathbb Z}_2$ gauge symmetries in these models can arise from solely
two-spin interactions. That one can generate effective multi-spin
interactions in some low energy limit of a two-spin Hamiltonian is not
unexpected; what is novel is that the symmetries we discuss are {\it
  exact}. We articulate a notion
of combinatorial gauge symmetry that underlies why it is possible to
construct local two-spin Hamiltonians with an exact ${\mathbb Z}_2$
gauge symmetry.


{\it Algebra-preserving transformations and monomial matrices} --
We start with a set of $N$ spin-1/2 degrees of freedom, such as the familiar spin models on a lattice with $N$ sites. The spin operators are Pauli matrices $\sigma^\alpha_i$, where $\alpha={\rm x,y,z}$ and $i=1,\dots, N$. Spins on different sites commute, while those on the same site satisfy the usual angular momentum algebra. Let us ask a simple question: which transformations of
these $3N$ operators can preserve all commutation and anti-commutation
relations? For $N$ bosons or fermions, this is a trivial question to answer; the allowed set of single-particle transformations belong to the unitary group $U(N)$ because either the commutation or anti-commutation relations need to be satisfied. But for spins, the question is harder; one cannot simply mix spatial components of different spins and retain both the intra- and inter-site algebra.

The Hilbert space for $N$ spins is $2^N$-dimensional and the allowed
operators in this space are $2^N\times 2^N$ unitary matrices,
corresponding to the group $SU(2^N)$. A generic transformation on
the spin operators, $\sigma_i^a \to U\,\sigma_i^a\,U^\dagger$
preserves the algebra, but also acts simultaneously on many spins: it
mixes the $3N$ single-spin operators $\sigma^a_i$ with the other
(multi-spin) $2^{2N}-1-3N$ generators of $SU(2^N)$. Therefore, if one
is to remain with only single-spin terms, one must work with a much
smaller subgroup of $SU(2^N)$. The simplest solution is trivial: only
rotate spins individually by restricting the allowed transformations
to $SU(2)\otimes SU(2)\otimes\cdots\otimes SU(2)$ or $N$ copies of
$SU(2)$. A more interesting and non-trivial solution is to also allow
permutations of spins. (If one wishes to connect to quantum gates,
these transformations correspond to the combination of one-qubit
rotations and the use of two-qubit SWAP gates.)

Any $SU(2)$ transformation on spin $i$ can be represented by a matrix in the rotation group $g_i\in SO(3)$ that acts on the spatial components of the vector $\vec{\sigma}_i=(\sigma^{\rm x}_i,\sigma^{\rm y}_i,\sigma^{\rm z}_i)^\top$. This representation makes it convenient to combine permutations and single-spin transformations into monomial matrices. Monomial matrices are generalizations of permutation
matrices such that the non-zero elements in each row and column are
group elements, not simply equal to 1. Here is an $N=4$ example:  
\begin{equation}
\begin{pmatrix}
\vec\sigma_1 \\
\vec\sigma_2 \\
\vec\sigma_3 \\
\vec\sigma_4 
\end{pmatrix}
\to
\begin{pmatrix}
0 & 0 & g_1 & 0 \\
0 & 0 & 0 & g_2 \\
g_3 & 0 & 0 & 0 \\
0 & g_4 & 0 & 0 
\end{pmatrix}\;
\begin{pmatrix}
\vec\sigma_1 \\
\vec\sigma_2 \\
\vec\sigma_3 \\
\vec\sigma_4 
\end{pmatrix}\;.
\label{eq:monomial}
\end{equation}
It is clear from this form that monomial matrices are orthogonal and that the product of any two monomial matrices is another monomial matrix. The example above can be written as a product of the diagonal matrix Diag($g_1,g_2,g_3,g_4$) and a $4\times 4$ permutation matrix.   

For arbitrary $N$, the group of monomial matrices is a semidirect product of the group generated by the diagonal matrices Diag($g_1,\cdots,g_N$) and the group of permutations (symmetric group) $S_N$. In mathematical literature, this particular form of a semidirect product is sometimes referred to as a wreath product.

To summarize the above: we are pointing out that many-body spin states admit a group of non-trivial transformations on the $3N$ spin components that preserve all spin algebras. When formulated in this way, the combination of local and permutation symmetry will allow us to construct exact lattice gauge theories using only two-body interactions.


{\it Combinatorial gauge symmetry} --
One particular subgroup of monomial transformations, such as in
Eq.~(\ref{eq:monomial}), is for $SO(3)$ rotations by angle $\pi$
around a given axis, which we take to be $\hat x$. This is equivalent
to flipping the $z$-component of spin. We shall use this special case
to construct a microscopic model with local ${\mathbb Z}_2$
symmetry. We term our methodology \textit{combinatorial gauge
  symmetry} for its relation to monomials and permutations.

\begin{figure}[!tbh]
\centering
\includegraphics[width=.35\textwidth]{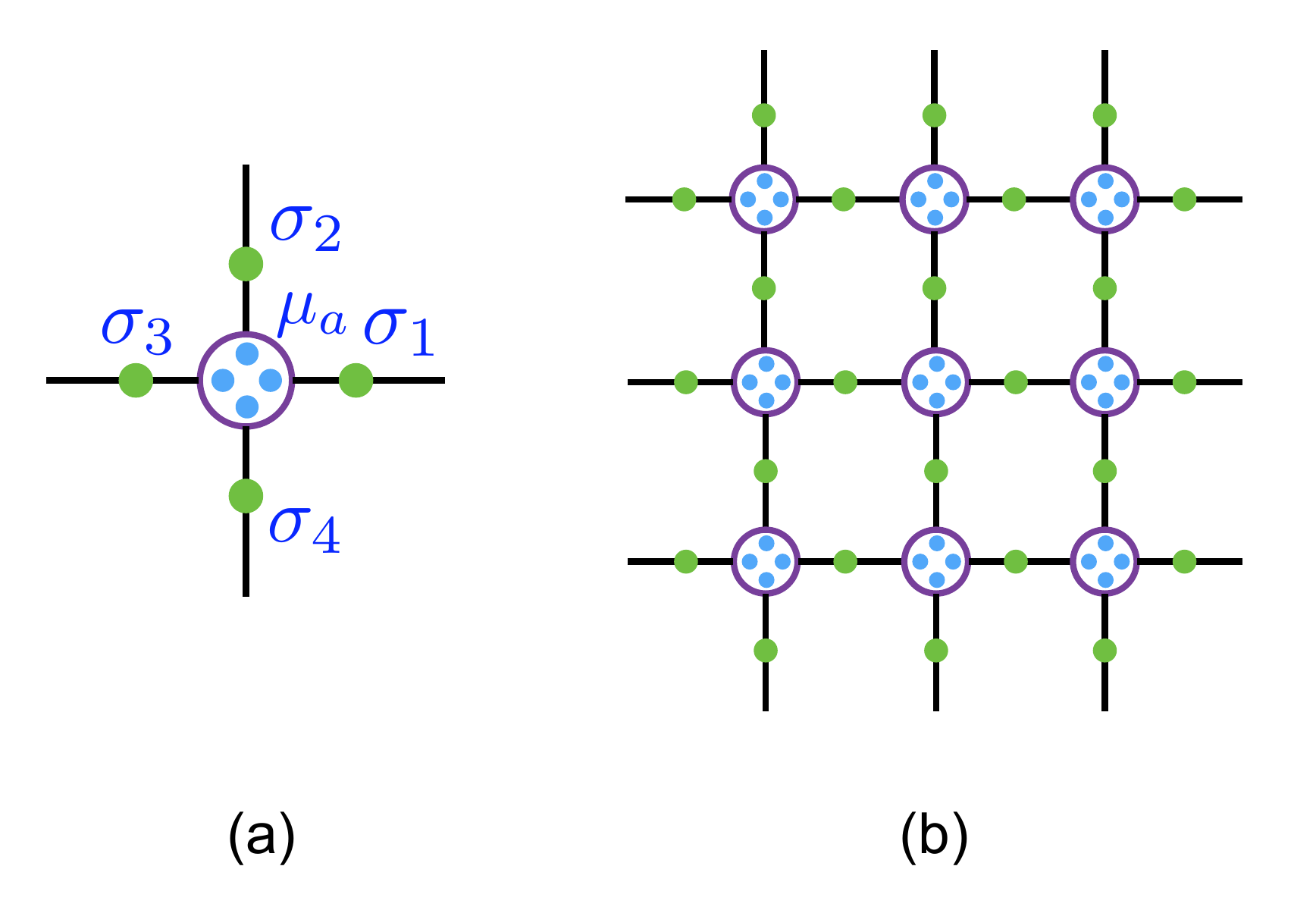}
\caption{(a) A single site (star) of the $\mathbb{Z}_2$ gauge theory, with 4 matter spins $\mu_a$ on the site, and 4 gauge spins $\sigma_i$ on the links. (b) The full lattice.}
\label{fig:star} 
\end{figure}

Consider the lattice depicted in Fig.~\ref{fig:star}, where 4
``matter" spins $\mu$ are placed on each lattice site, and ``gauge"
spins $\sigma$ are placed on the links. A single site (star) is
isolated in Fig.~\ref{fig:star}(a), and contains the 4 matter spins
and 4 gauge spins sitting on the links. The gauge spins are shared by
neighboring stars, as depicted in Fig.~\ref{fig:star}(b). Each matter
spin couples only to its neighboring gauge spins but not to one
another (or other lattice sites). Gauge spins do not couple to each
other. We encode all two-spin ($ZZ$) couplings between $\mz_a$ and
$\sz_i$ by a $4\times 4$ matrix $W_{ai}$.

The quantum fluctuations will come from two transverse fields $\widetilde\Gamma$ and $\Gamma$ acting on the gauge spins and matter spins, respectively. For generality, we allow $\Gamma$ and $\widetilde\Gamma$ to have different magnitudes.  

Thus the full lattice Hamiltonian is given by
\begin{align}
H=-\sum_s
\left[
  J\sum_{\substack{a\in s\\i\in s}} \,W^{}_{ai}\;\sz_i\,\mz_a
  +\Gamma \,\sum_{a\in s} \mx_a
  \right]-\widetilde\Gamma \sum_i \sx_i
\;,
\label{eq:toric_code_lattice_exact}
\end{align}
where the $s$ are stars on the lattice. 

We shall select the interaction matrix $W$ so as to satisfy the
monomial transformations as in Eq.~(\ref{eq:monomial}) that act on the
$z$-components of the gauge and matter spins as follows:
  \begin{align}
\sz_i&\rightarrow \sum_{j=1}^4 R_{ij}\;\sz_j
\nonumber\\
\mz_a&\rightarrow \sum_{b=1}^4 \mz_b\;(L^{-1})_{ba}
\;.
\label{eqs:RLtransform}
  \end{align}
These are monomial transformations that preserve the spin commutation
and anticommutation relations, as discussed above. The $L$ (``left")
and $R$ (``right") matrices act like gauge transformations on the
$z$-components of the gauge and matter spins. These monomial matrices
have elements $\pm 1$. (Henceforth all monomial matrices will be of
this kind.)

The requirement that the Hamiltonian
Eq.~(\ref{eq:toric_code_lattice_exact}) be invariant with respect to
transformations Eq.~(\ref{eqs:RLtransform}) is equivalent to the
requirement that the $W$ matrices be invariant under the automorphism
transformation $L^{-1}WR=W$, where $L$ and $R$ are $4\times4$ monomial
matrices~\cite{Automorphism}. [The transverse fields are also invariant under
  the transformation Eq.~(\ref{eqs:RLtransform}), and we shall return
  to this point below.]

Hadamard matrices~\cite{Automorphism} satisfy these conditions. These
matrices have elements $\pm 1$, and all its columns (or rows) are
orthogonal vectors, i.e., $W^\top\,W\propto \openone$. (They maximize
the determinant of the information matrix $W^\top\,W$.) We pick an
intuitive form of $W$, where the coupling between $\sz_i$ and $\mz_a$
is anti-ferromagnetic when $i=a$ and ferromagnetic otherwise:
\begin{equation}
W=
\begin{pmatrix}
-1 & +1 & +1 & +1 \\
+1 & -1 & +1 & +1 \\
+1 & +1 & -1 & +1 \\
+1 & +1 & +1 & -1 
\end{pmatrix}.
\label{eq:W}
\end{equation}
All other choices of $W$ are equivalent by symmetry and will not
affect the spectrum. Specifically, any two Hadamard matrices $W$ and
$W^\prime$ are equivalent if there exist monomial matrices $S_1,S_2$
such that $W^\prime=S_1^{-1}WS_2$.

Our model further restricts $R$ to be diagonal because any
off-diagonal permutation of gauge spins would deform the lattice. For
example, with our choice of $W$ in Eq.~(\ref{eq:W}), the following
pair satisfies the conditions above:
\begin{align}
L=&
\begin{pmatrix}
0 & +1 & 0 & 0 \\
+1 & 0 & 0 & 0 \\
0 & 0 & 0 & -1 \\
0 & 0 & -1 & 0 
\end{pmatrix}
&
R=&
\begin{pmatrix}
-1 & 0 & 0 & 0 \\
0 & -1 & 0 & 0 \\
0 & 0 & +1 & 0 \\
0 & 0 & 0 & +1 
\end{pmatrix}
\label{eq:LR}
\;.
\end{align}
Once we choose an $R$, we determine $L$ uniquely by solving the automorphism condition: $L=W\,R\,W^{-1}$. (The number of $-1$'s in the diagonal $R$ matrix must be even so that the corresponding $L$ is a monomial matrices.) Note that flipping gauge spins, even without permuting them, requires a simultaneous permutation of matter spins.

\begin{figure}[!t]
\centering
\includegraphics[width=.45\textwidth]{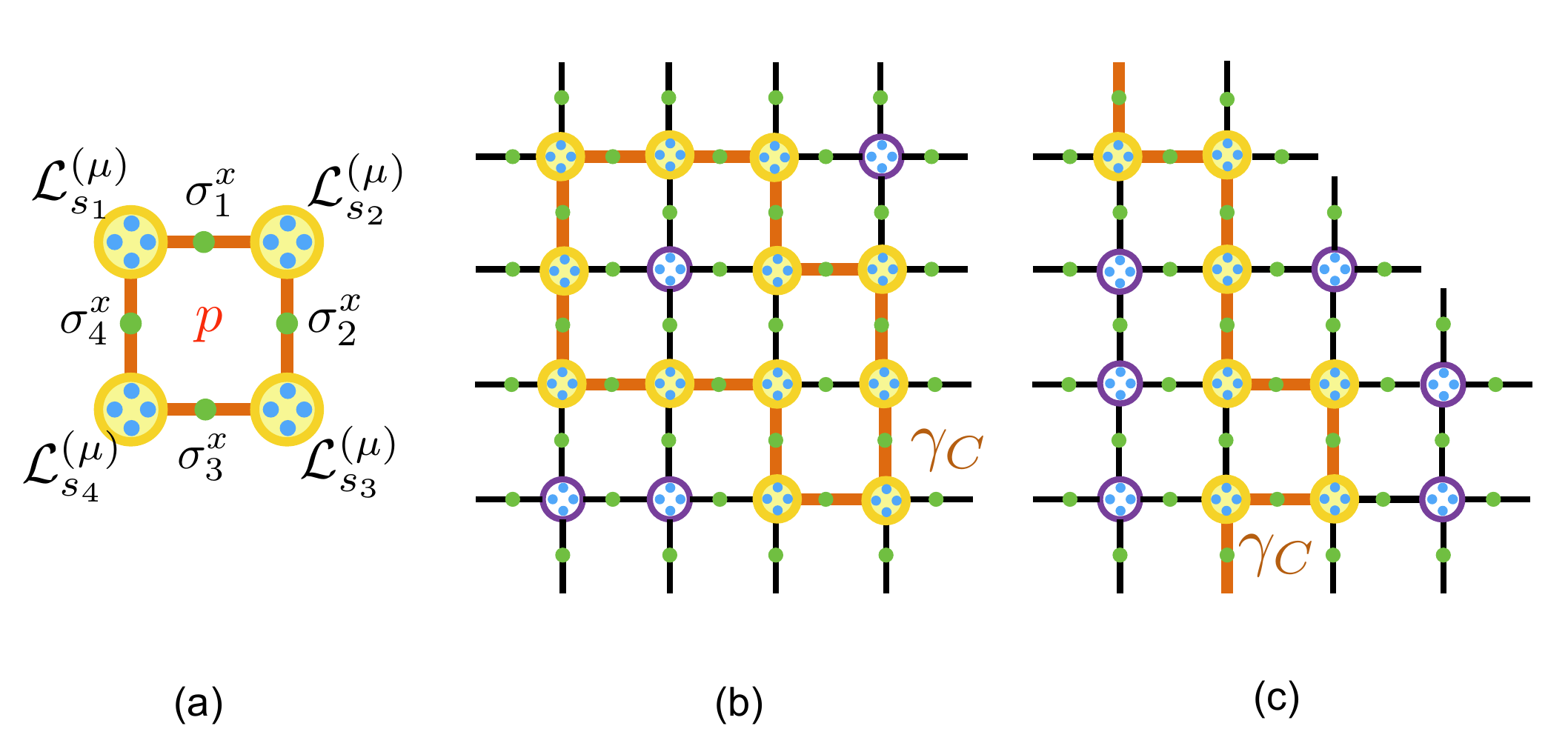}
\caption{(a) Operator generating the local $\mathbb{Z}_2$ gauge transformation on an elementary plaquette, $G_p$ in Eq.~(\ref{eqs:G_p}) (b) A closed loop operator along a path $\gamma_C$. (c) An open string operator along a path in a system with boundaries.}
\label{fig:paths} 
\end{figure}

The automorphism pair $(L, R)$ directly leads to the local $\mathbb{Z}_2$ gauge symmetry of the full lattice Hamiltonian~(\ref{eq:toric_code_lattice_exact}). Consider an elementary plaquette $p$ depicted in Fig.~\ref{fig:paths}(a) and define the local gauge transformation 
\begin{align}
G_p = \prod_{s \in p} \mathcal{L}^{(\mu)}_s\prod_{s \in p} \mathcal{R}_s^{(\sigma)},
\label{eqs:G_p}
\end{align}
where $\mathcal{L}^{(\mu)}_s$ denotes the operator that permutes and flips the matter spins at each corner site $s$ of the plaquette as in Eq.~(\ref{eq:LR}): 
$\mathcal{L}^{(\mu)}_s\mz_a\;(\mathcal{L}^{(\mu)}_s)^{-1}
    =
    \sum_b \mz_a\;(L^{-1})_{ba}$, as in the transformation in Eq.~(\ref {eqs:RLtransform}), and similarly for $\mathcal{R}_s^{(\sigma)}$ on the gauge spins: $\mathcal{R}_s^{(\sigma)} \sz_i \ (\mathcal{R}_s^{(\sigma)})^{-1} = \sum_j R_{ij} \sz_j$.
$\mathcal{L}^{(\mu)}_s$ is uniquely determined by the local operator $\mathcal{R}^{(\sigma)}_s$ that flips the two gauge spins on links emanating from the site $s$ -- just as $L$ is determined by $R$. Since here we restrict $R$ to be diagonal, corresponding to only flipping the gauge spins without permuting them, the spin flip $\sz \rightarrow -\sz$ is simply generated by $\sx$. Therefore we have $\prod_{s\in p} \mathcal{R}_s^{(\sigma)} = \prod_{i\in p} \sx_i$, where $i$ runs over all gauge spins in a plaquette.
Any two $L$ matrices commute and therefore the plaquette operators do as well, $[G_p,G_{p^\prime}]=0$. 

The importance of $G_p$ is that it is a local symmetry of the full lattice Hamiltonian~(\ref{eq:toric_code_lattice_exact}): $[H, G_p]=0$, for all $p$. Invariance of the Ising interaction term follows from the automorphism above, while invariance of the transverse field terms $\Gamma$ and $\widetilde\Gamma$ follows from two observations. First, all spin flips by the operator pair $(\mathcal{L}^{(\mu)}_s,\mathcal{R}^{(\sigma)}_s)$ can be viewed as $180^\circ$ rotations around the $x$-axis, which commute with $\sx$ and $\mx$. Second, the transverse fields are uniform and therefore independent of permutations. Therefore, the Hamiltonian~(\ref{eq:toric_code_lattice_exact}) is a gauge theory with a local $\mathbb{Z}_2$ gauge symmetry that is generated by $G_p$. This symmetry relies on the locking of the permutations contained in the operators $\mathcal{L}^{(\mu)}_s$ to the $\mathbb{Z}_2$ transformations in the $\mathcal{R}^{(\sigma)}_s$, which is another reason that we refer to it as combinatorial gauge symmetry.

One can further construct loop or closed string symmetry operators on the lattice, as shown in Fig.~\ref{fig:paths}(b). For systems with boundaries, one can also associate a symmetry operation to open strings, as depicted in Fig.~\ref{fig:paths}(c). The loop (or string) operator along a path is composed of both the gauge spin flips $\prod_\ell \sx_\ell$, where $\ell$ are the links along the path, as well as the corresponding operations on matter spins $\prod_s\mathcal{L}^{(\mu)}_s$ applied to each star along the path. In the case of closed paths, the loop operator is equivalent to a product of all plaquette operators $G_p$ enclosed by the loop.

{\it Special case: ${\mathbb Z}_2$ gauge theory} --
The Hamiltonian Eq.~(\ref{eq:toric_code_lattice_exact}) obeys a local
${\mathbb Z}_2$ gauge symmetry for \textit{all} values of the
parameters $J, \ \Gamma$, and $\widetilde \Gamma$. Here we shall
obtain, as a particular limit, an effective Hamiltonian with a 4-spin
interaction on a star, which lands directly onto the more familiar
${\mathbb Z}_2$ gauge theory on the square
lattice~\cite{z2_Wegner,z2_Kogut}, in the following manner.

Isolate a single star with its 4 spins $\mu$ on the site and 4 gauge
spins $\sigma$ on the links, as depicted in
Fig.~\ref{fig:star}(a). Let us freeze for the moment a given
configuration of the gauge spins $\sz_i, i=1,2,3,4$ in the
$z$-basis. The Hamiltonian~(\ref{eq:toric_code_lattice_exact}) for
each matter spin $\mu_a$ on a star can be viewed as that of a single
spin in a magnetic field, whose eigenvalues are functions of $\sz_i$:
\begin{align}
E^{\,(\pm)}_a(\sz_1,\sz_2,\sz_3,\sz_4)=\pm \left[
J^2\,\left(\sum_{i=1}^4\;W^{}_{ai}\;\sz_i\right)^2
+
\Gamma^2\right]^{1/2}
\label{eq:eigenvalues_sqrt}
\;.
\end{align}
The expression in Eq.~(\ref{eq:eigenvalues_sqrt}) can be written, for
any value of $\Gamma$ and $J$, as
\begin{align}
&
E_a^{\,(\pm)}=\pm C_0
\pm C_2\;\sum_{i\ne j}^4 W^{}_{ai}\,W^{}_{aj}\;\sz_i\,\sz_j
\nonumber\\
&\qquad
\pm C_4\;W^{}_{a1}\,W^{}_{a2}\,W^{}_{a3}\,W^{}_{a4}\;
\sz_1\,\sz_2\,\sz_3\,\sz_4
\label{eq:eigenvalues_sqrt_expand}
\;,
\end{align}
where $C_0$,~$C_2$ and $C_4$ are constants that depend on $J$ and
$\Gamma$. This expression follows from expanding the square root in
Eq.~(\ref{eq:eigenvalues_sqrt}) in powers of the $\sz_i$ and using
$(\sz_i)^2=1$ and $(W^{}_{ai})^2=1$; the binary polynomial inside the
square root terminates and the only terms that remain are of the form
in Eq.~(\ref{eq:eigenvalues_sqrt_expand}). While the expansion is
useful in proving the identity between
Eqs.~(\ref{eq:eigenvalues_sqrt}) and
(\ref{eq:eigenvalues_sqrt_expand}), we remark that the result is exact
(non-perturbative), because both expressions only take values in
discrete sets.

The low energy manifold of states corresponds to the sum over the lowest eigenvalues, $H_{\rm eff}^{\rm
  star} = \sum_{a=1}^4 E_a^{\,(-)}$, which is separated from the next
levels by a gap of size at least $2|\Gamma|$. We thus arrive at the
following simple effective Hamiltonian for a single star:
\begin{align}
H_{\rm eff}^{\rm star}
=
\gamma - \lambda \;\sz_1\,\sz_2\,\sz_3\,\sz_4
\;,
\label{eq:star}
\end{align}
where the coefficients $\gamma$ and $\lambda$ are functions of
$\Gamma$ and $J$ are explicitly given in the Supplemental
Material. These relations follow from the consistency between
Eqs.~(\ref{eq:eigenvalues_sqrt}) and
(\ref{eq:eigenvalues_sqrt_expand}). The parity $P\equiv
\sz_1\,\sz_2\,\sz_3\,\sz_4$ for the ground state of
Eq.~(\ref{eq:star}) is $P=+1$, since $\lambda>0$. By modifying the
matrix $W$, we could flip the sign of $\lambda$ and have instead the
$P=-1$ parity sector as the ground state (for example, by flipping the
sign of any one column of $W$).

Let us now turn to the low energy effective model for the whole lattice. In the limit $|\Gamma| \gg J$, we find the effective Hamiltonian
\begin{align}
H_{\rm eff} = - \lambda \sum_s \;\prod_{i\in s} \sz_i
-\widetilde \Gamma \sum_i \sx_i \;.
\label{eq:toric_code_lattice}
\end{align}
This Hamiltonian is exactly that of the $\mathbb{Z}_2$ quantum gauge
theory, which supports a topological phase for
$\widetilde\Gamma/\lambda$ below a threshold. To get the toric/surface
code limit, one only has to notice that the lowest order term that
survives in a perturbation theory in $\widetilde\Gamma/\lambda$ is the
term that flips all spins around a plaquette~\cite{PhysRevD.17.2637,
  fradkin2013field, subir}.

Taking $|\Gamma|\to \infty$, while keeping $\lambda$ fixed, opens an
infinite gap to the excited sectors, where at least one
$E_a^{(-)}$ becomes $E_a^{(+)}$. The splitting $2|\lambda|$ between the two parity states
within the lowest energy sector remains finite. The expansion of $\lambda$ in the regime of $J\ll\Gamma$ yields
$\lambda = 12J^4/\Gamma^3 + \mathcal{O}(J^6/\Gamma^5)$. (Note that
terms of order $\Gamma$ vanish.) To access this regime we would fix
$\lambda$ and tune $J$ such that
$J=\left|\lambda\,\Gamma^3/12\right|^{1/4}$.
Physically, in this limit the matter fields $\mu$ can be ``integrated
out'' to obtain the exact four-spin effective Hamiltonian.

We corroborate the above analytical features with numerical studies in
the Supplemental Material. All degeneracies are confirmed to machine
precision.

\begin{figure}[!t]
\centering
\includegraphics[width=.45\textwidth]{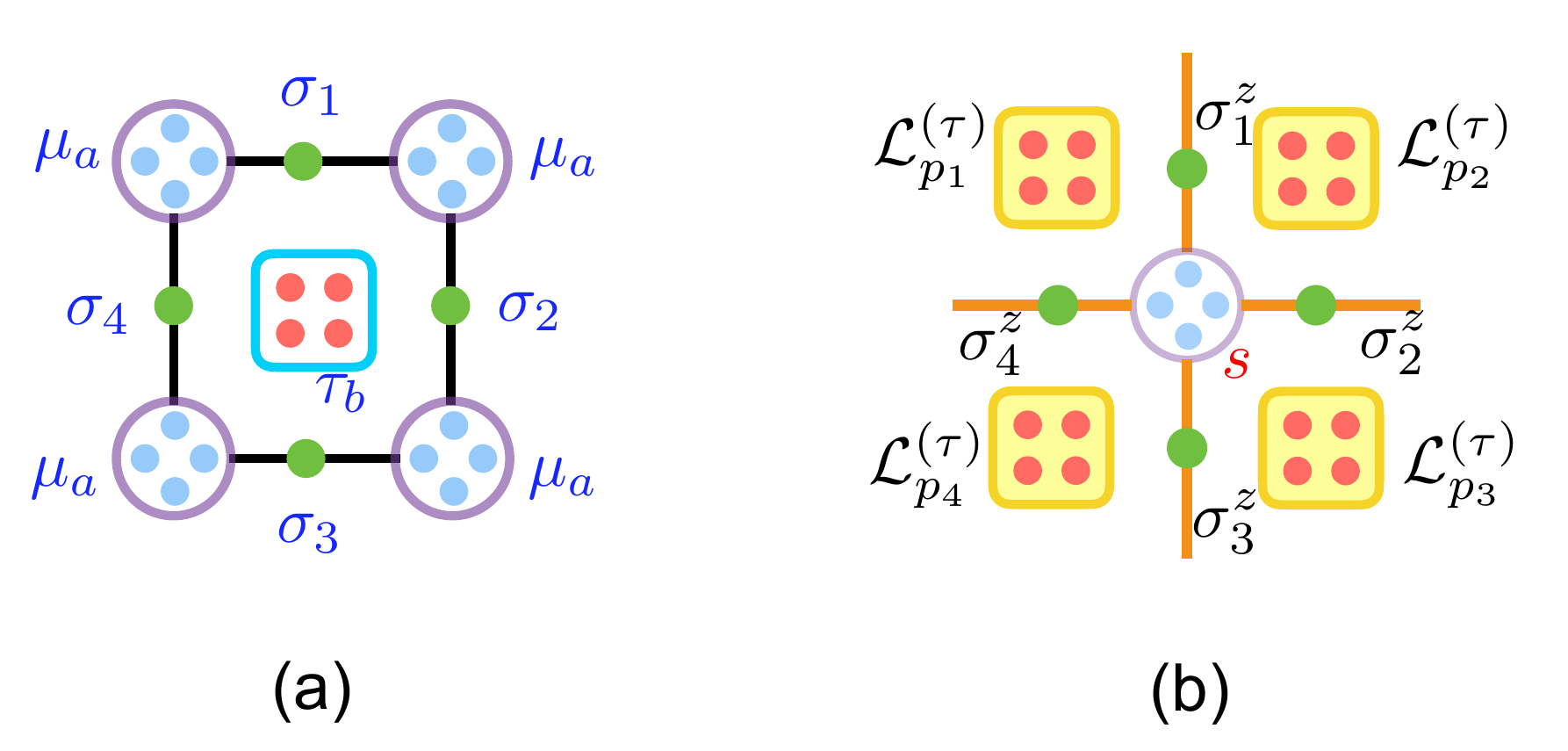}
\caption{(a) A single plaquette of the $\mathbb{Z}_2$ gauge theory, with 4 gauge spins $\sigma_i$ on the links, 4 matter spins $\mu_a$ on the site,  and 4 additional matter spins $\tau_b$ at the center of the plaquette. (b) A single star operator $F_s$.}
\label{fig:dual} 
\end{figure}


{\it Combinatorial gauge symmetry for both electric and magnetic
  loops} -- So far we used the combinatorial gauge symmetry to
construct a model with $\mathbb{Z}_2$ plaquette operators. Here we
shall construct a model with both $\mathbb{Z}_2$ plaquette and star
operators, just as in the toric code, but still using {\it only} at
most two-body interactions.

We add another four spin-1/2 degrees of freedom to the
center of all plaquettes in addition to the ones on the star. We
denote these additional matter spins on the dual lattice as $\tau$,
shown in Fig.~\ref{fig:dual}(a). Furthermore, the pairwise interaction
couples $\tx$ and $\sx$, i.e. an $XX$ interaction. The full
Hamiltonian is:
\begin{align}
H=
&-\sum_s
\left[
  J\sum_{\substack{a\in s\\i\in s}} \,W^{}_{ai}\;\sz_i\,\mz_a
  +\Gamma \,\sum_{a\in s} \mx_a
  \right]
\nonumber\\
&-\sum_p
\left[
  J\sum_{\substack{b\in p\\j\in p}} \,W^{}_{bj}\;\sx_j\,\tx_b
  +\Gamma \,\sum_{b\in p} \tz_b\;
  \right]
\;.
\label{eq:exact_plaquette_and_star}
\end{align}
In other words, on the dual lattice the spin components are
transformed by $X\leftrightarrow Z$ relative to the original
lattice. There is no need for a transverse field on the gauge $\sigma$
spins in this model; quantum dynamics is already present through the
presence of both $XX$ and $ZZ$ interactions.

By analogy with the plaquette operators $G_p$ in
Eq.~(\ref{eqs:G_p}), there is a set of star operators $F_s$, according to combinatorial gauge symmetry, which
exist on the dual lattice [see Fig.~\ref{fig:dual}(b)]:
\begin{align}
F_s = \prod_{p \in s} \mathcal{L}^{(\tau)}_p\prod_{i\in s} \sz_i~.
\label{eqs:F_s}
\end{align}
The dual, ``left" operators $\mathcal{L}^{(\tau)}_p$ flip $\tau$ spins
in the $x$-basis just like the operators $\mathcal{L}^{(\mu)}_s$ in
Eq.~(\ref{eqs:G_p}) flip $\mu$ spins in the $z$-basis. By
construction, these two operators commute:
$[\mathcal{L}^{(\mu)}_s,\mathcal{L}^{(\tau)}_p]=0$. Therefore, we have
a star and a plaquette operator that also commute: $[G_p,F_s]=0$,
exactly as in the toric code. It is easy to check that the Hamiltonian
commutes with both stars and plaquettes: $[H,G_p]=[H,F_s]=0$.

Given the commuting set of star and plaquette operators, the
Hamiltonian in Eq.~(\ref{eq:exact_plaquette_and_star}) is equivalent
to the toric code in the asymptotic limit of large $\Gamma$,
except that it contains \textit{only} two-body interactions
and fields. This is a direct result of the combinatorial gauge
symmetry.

{\it Extension to other topological states} --
Fracton topological
phases~\cite{PhysRevLett.94.040402,BRAVYI2011839,PhysRevA.83.042330,
  PhysRevB.94.235157} (for a review, see
Ref. ~\onlinecite{nandkishore2019fractons}) are novel phases of matter
with a robust sub-extensive ground state degeneracy and with
excitations that are strictly immobile, or constrained to move within
a subdimensional manifold. Apart from theoretical interest such as
classifications of phases of matter and formulations in terms of
higher-rank gauge theories~\cite{PhysRevB.95.115139}, fracton systems
are also believed to hold promise for fault-tolerant quantum
computation, as well as robust quantum
memory~\cite{PhysRevA.83.042330}. In spite of the intensive
theoretical investigations on fractonic models, experimental
realizations directly in terms of spins have barely been
discussed~\cite{you2018majorana}.

The building blocks of our $\mathbb{Z}_2$ gauge theory can also be
used to construct 3D models, such as one of the simplest fractonic
model, the X-cube~\cite{PhysRevB.81.184303, PhysRevB.94.235157}. The
construction with matter and gauge spins parallels closely that in 2D,
and we provide details for the construction of both the 3D toric code
and the X-cube model in the Supplemental Material.

{\it Summary and outlook} -- We have argued that many-body spin states
admit a combinatorial gauge symmetry and we have used it to construct
quantum spin liquids out of only two-body and single-body terms. The
symmetry holds exactly for all ranges of parameters in the
Hamiltonians that we have constructed. This presents an alternative
path to explore quantum spin liquids in systems without four-body (or
higher) interaction terms. Our approach may prove useful in the quest
for topological qubits (via surface codes), which can potentially be
hosted by certain quantum spin liquids.

{\it Acknowledgments} -- 
The work by C.~C. and Z.-C.~Y. is supported by the DOE. The part of
  the work centered on topological phases of matter is supported by
  DOE Grant No. DE-FG02-06ER46316; the part of the work centered on
  quantum information science is supported by Grant No. DE-SC0019275.

\bibliography{reference}

\newpage
\onecolumngrid
\appendix
\section*{Supplemental Material for "Constructing Quantum Spin Liquids Using Combinatorial Gauge Symmetry"}

In this Supplemental Material, we discuss: (1) large $J$ limit of Hamiltonian Eq.~(\ref{eq:toric_code_lattice_exact}) in the main text; (2) numerical studies of the gauge-matter Hamiltonian; (3) construction of 3D toric code; and (4) construction of the X-cube model.

\subsection{Large $J$ limit of Hamiltonian Eq.~(\ref{eq:toric_code_lattice_exact})}
The classical limit of Hamiltonian~(\ref{eq:toric_code_lattice_exact}) has finite $J$ and no transverse fields $\Gamma=\widetilde\Gamma=0$. It is straightforward to verify that the ground state of each star on the lattice is eightfold degenerate with energy $-8J$ and parity $P=+1$. The eight ground state configurations of the gauge and matter spins on each star are shown in Table~\ref{table:ClassicalDegeneracy}. 
\begin{table}[h]
\begin{tabular}{c c c c | c c c c }
\hline
$\sz_1$ & $\sz_2$ & $\sz_3$ & $\sz_4$ & $\mz_1$ & $\mz_2$ & $\mz_3$ & $\mz_4$ \\
\hline
$\ua$ & $\ua$ & $\ua$ & $\ua$ & $\da$ & $\da$ & $\da$ & $\da$ \\
$\ua$ & $\ua$ & $\da$ & $\da$ & $\ua$ & $\ua$ & $\da$ & $\da$ \\
$\ua$ & $\da$ & $\ua$ & $\da$ & $\ua$ & $\da$ & $\ua$ & $\da$ \\
$\ua$ & $\da$ & $\da$ & $\ua$ & $\ua$ & $\da$ & $\da$ & $\ua$ \\
\hline
$\da$ & $\da$ & $\ua$ & $\ua$ & $\da$ & $\da$ & $\ua$ & $\ua$ \\
$\da$ & $\ua$ & $\da$ & $\ua$ & $\da$ & $\ua$ & $\da$ & $\ua$ \\
$\da$ & $\ua$ & $\ua$ & $\da$ & $\da$ & $\ua$ & $\ua$ & $\da$ \\
$\da$ & $\da$ & $\da$ & $\da$ & $\ua$ & $\ua$ & $\ua$ & $\ua$ \\
\hline
\end{tabular}
\caption{Eight ground states in the classical limit $J\gg\Gamma,\widetilde\Gamma$}
\label{table:ClassicalDegeneracy}
\end{table}
The energy gap is $4J$. Note that the matter spins are ``slaved" to the gauge spins. 

Now apply small transverse fields $\Gamma\ll J$ and $\widetilde\Gamma\ll J$ to both the matter and gauge spins. In degenerate perturbation theory the lowest order term that restores the system to the classical ground state manifold is the plaquette operator $G_p$. It flips four gauge spins around a plaquette plus the corresponding eight matter spins at the corners, leaving the ground state manifold degenerate. The Hamiltonian~(\ref{eq:toric_code_lattice_exact}) in this limit is composed of a star and plaquette term:
\begin{align}
H=-\sum_s
\left[
  J\sum_{\substack{a\in s\\i\in s}} \,W^{}_{ai}\;\sz_i\,\mz_a
  \right]+g\sum_p G_p
\;.
\label{eq:toric_code_classical}
\end{align}
$g$ is the energy scale of the plaquette term on the order of ($\Gamma^8\widetilde\Gamma^4)/J^{11}$, corresponding to the number of spin flips to lowest order.  Because the local gauge symmetry is preserved at all stages, a quantum spin liquid state should emerge in this limit. Again, it is made possible by the combinatorial symmetry which ensures that no spin ordered state is favored. The Hamiltonian~(\ref{eq:toric_code_classical}) is reminiscent of the well-known toric code with its star and plaquette terms~\cite{tqc_Kitaev}. The toric code also emerges as a limit of the conventional $\mathbb{Z}_2$ lattice gauge theory to lowest order of perturbation theory~\cite{PhysRevD.17.2637,
  fradkin2013field, subir}.

\subsection*{Numerical study of the gauge-matter Hamiltonian Eq.~(\ref{eq:toric_code_lattice_exact})}
\label{sec:full_lattice_numerics}

Here we provide a numerical comparison of the effective Hamiltonian
with the four-spin interaction and the Hamiltonian with the matter and
gauge spins. The studies focused on clusters containing either a
single star or a single plaquette surrounded by a fixed background, in
the presence of the transverse field $\widetilde\Gamma$. We have shown in the main text that the gauge-matter Hamiltonian for a single star in the absence of $\widetilde{\Gamma}$
\begin{equation}
H^{\rm star}= -J\sum_{\substack{a\in s\\i\in s}} \,W^{}_{ai}\;\sz_i\,\mz_a
  -\Gamma \,\sum_{a\in s} \mx_a
\label{eq:star_full}
\end{equation}
corresponds to the effective Hamiltonian~(\ref{eq:star}) at low energy, with:
\begin{align}
\gamma &= -\frac{1}{2}\left(\sqrt{\Gamma^2+16J^2} +3|\Gamma|+4\sqrt{\Gamma^2+4J^2}\right)
\nonumber\\
\lambda &= -\frac{1}{2} \left(\sqrt{\Gamma^2+16J^2}+3|\Gamma|-4\sqrt{\Gamma^2+4J^2}\right)
\;.
\label{eqs:gamma-lambda}
\end{align}

The
analysis passes each of the following three tests in favor of a spin
liquid \textit{within machine precision}.

\begin{figure*}
\centering
\includegraphics[width=.9\textwidth]{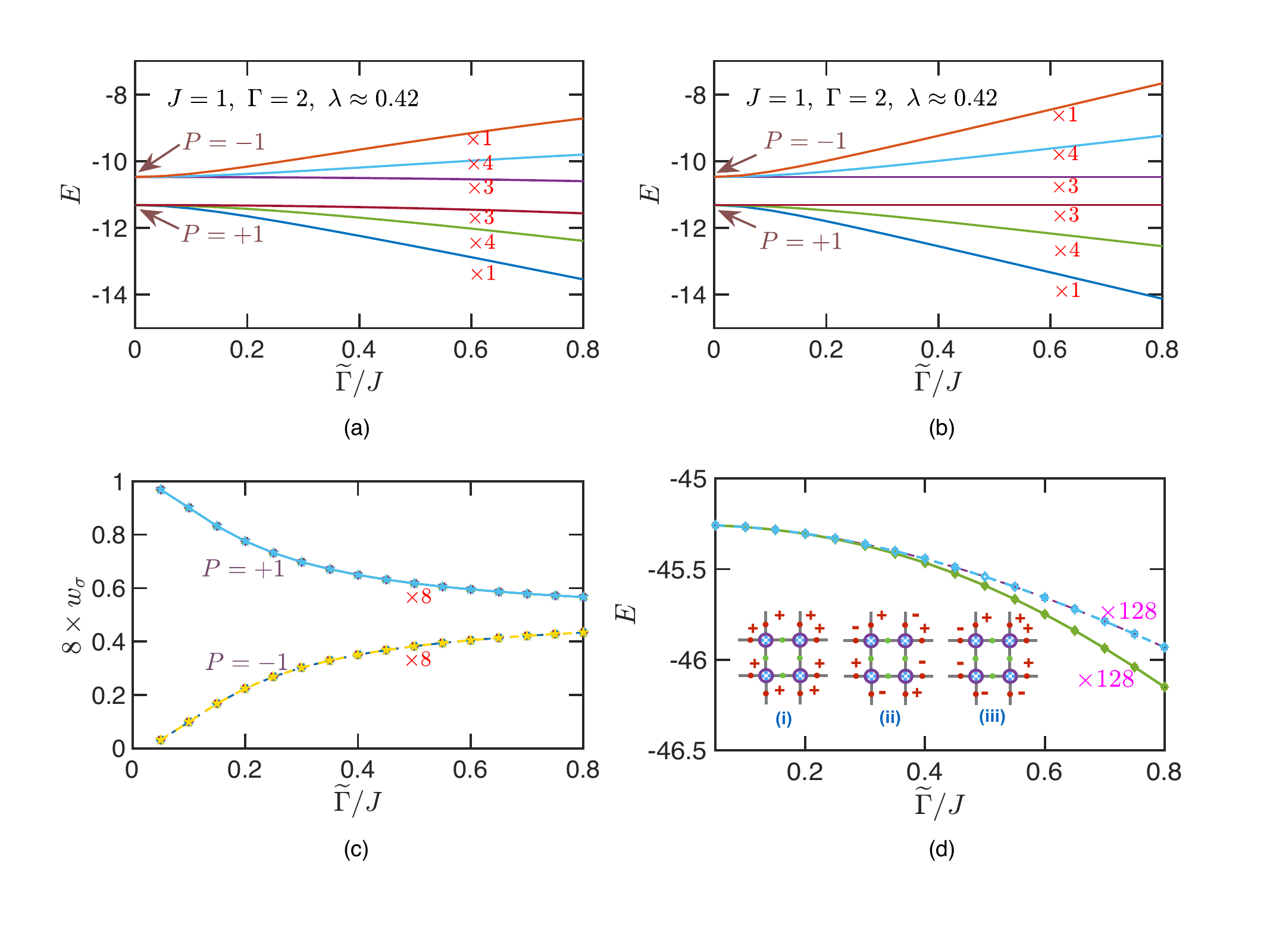}
\caption{(a)\&(b) Spectrum of a single star as a function of the transverse field $\widetilde{\Gamma}$. (a) Lowest energy sector of the full Hamiltonian~(\ref{eq:star_full}) plus a transverse field $\widetilde{\Gamma}$ on the gauge spins. (b) Complete spectrum of the effective Hamiltonian~(\ref{eq:star}) plus a transverse field $\widetilde{\Gamma}$ on the gauge spins. (c) Weight of the ground state wavefunction of Hamiltonian~(\ref{eq:star_full}) in the presence of a transverse field $\widetilde{\Gamma}$ on each of the eight $P=+1$ and eight $P=-1$ configurations of the star operator, as a function of $\widetilde{\Gamma}$. The eight curves in each set fall on top of one another. (d) The ground state and the first excited state energies corresponding to the external leg configurations. The (i)-(iii) insets are examples. For all panels, the degeneracy of each energy level, or the number of curves on top of one another is labeled below the curve by $\times$. We choose $J=1$ and $\Gamma=2$, yielding an excitation gap between the two parity sectors $2\lambda \approx 0.84$.}
\label{fig:toric_combined} 
\end{figure*}

(i) Fig.~\ref{fig:toric_combined}(a)\&(b): For a single star in the
presence of the $\widetilde\Gamma$ field on the gauge spins, the
eigenstates of Eq.~(\ref{eq:star}) supplemented by the transverse
field on the gauge spins have a one-to-one correspondence to the
lowest energy sector levels of the Hamiltonian~(\ref{eq:star_full})
supplemented by the transverse field on the gauge spins, with
precisely the same degeneracies. This implies that the presence of
matter spins in the Hamiltonian~(\ref{eq:star_full}) does \textit{not}
break the parity symmetry among the gauge spins even with a
\textit{nonzero} $\widetilde{\Gamma}$.

(ii) Fig.~\ref{fig:toric_combined}(c): In the presence of $\widetilde\Gamma$,
the ground state wavefunction of the full
Hamiltonian~(\ref{eq:star_full}) is an equal weight superposition of
the eight $P=+1$ configurations and, with smaller but equal weights, the eight
$P=-1$ configurations, as should be the case for the exact star
operator in the effective Hamiltonian~(\ref{eq:star}). This indicates
that, at the single star level, our ``molecule''
Hamiltonian~(\ref{eq:star_full}) replicates the quantum dynamics of an
exact star operator, even when quantum fluctuations on both the matter
and gauge spins are present.

(iii) Fig.~\ref{fig:toric_combined}(d): the
ground state energy of a single \textit{plaquette}, as well as the gap
to the first excited state, is identical for any fixed configuration
of external legs surrounding the plaquette. We have confirmed that this holds for all 128
configurations of external legs, implying that the transverse field $\widetilde{\Gamma}$ does not
favor any ordering pattern on the lattice. Moreover, the gap
to the first excited state is independent of the environment,
meaning that the effective plaquette operator generated by
$\widetilde{\Gamma}$ does not favor any one spin order either.

The spectrum of the single star of the effective Hamiltonian
Eq.~(\ref{eq:star}) in the presence of a transverse field can be
obtained straightforwardly, leading to the following 16 eigenvalues
(degeneracies): $\pm\sqrt{\lambda^2+16\widetilde{\Gamma}^2}~(\times
1),~\pm\sqrt{\lambda^2+4\widetilde{\Gamma}^2}~(\times
4),~\pm\lambda~(\times 3)$. The constant shift $\gamma$ is not shown
for simplicity. The lowest eigenstate is an equal superposition of all
8 spin configurations that satisfy $P=+1$ and, with smaller amplitude
that varies with increasing $\widetilde\Gamma$, an equal superposition
of all 8 spin configurations that satisfy $P=-1$. Symmetry of the
ground state of the molecule is necessary, but not sufficient, for
symmetry in the lattice.

Diagonalization of the Hamiltonian Eq.~(\ref{eq:star_full}) in the
presence of the transverse field $\widetilde\Gamma$ yields states with
the exact same degeneracies as those of the effective model on a
transverse field presented above. Fig.~\ref{fig:toric_combined}(a) and
(b) present the spectrum for both the effective
Hamiltonian with the four-spin interaction and the lowest energy sector
for the Hamiltonian with the matter and gauge spins.

To compare the weights of the ground state wavefunction on each of the
gauge spin configurations, we take the ground state $|\psi\rangle$ of the Hamiltonian with gauge and matter spins and obtain the reduced density matrix by tracing over the matter spins: $\rho_{\bm \sigma} = {\rm tr}_{\bm \mu} |\psi\rangle \langle \psi|$, where ${\bm \mu}$ and ${\bm \sigma}$ stand for the four matter and gauge spins, respectively. Then we compute the weight of $\rho_{\bm \sigma}$ on each of the sixteen (eight with  $P=+1$ and eight with $P=-1$) configurations $|{\bm \sigma}_\ell\rangle$, $\ell=1, 2, \ldots, 16$:
\begin{equation}
{w}_{{\bm \sigma}_\ell} := {\rm tr} \left(\rho_{\bm \sigma} |{\bm \sigma}_\ell\rangle \langle {\bm \sigma}_\ell|\right).
\end{equation}
These weights are displayed in Fig.~\ref{fig:toric_combined}(c). The data show
that the ground state wavefunction contains an equal amplitude
superposition of the eight $P=+1$ gauge spin configurations and, with
smaller weight, the eight $P=-1$ configurations, mirroring exactly the
case for the star operator in the effective Hamiltonian with the
four-spin interaction. This indicates that, at the single star level, our
``molecule'' Hamiltonian~(\ref{eq:star_full}) replicates the quantum
dynamics of an exact star operator Eq.~(\ref{eq:star}), even when
quantum fluctuations on both the matter and gauge spins are present
via the transverse fields.

To further support our claim that the system does not favor any
ordered state when placed on the full lattice, we move to the next
level of complexity and focus on a single plaquette. Consider a single
plaquette surrounded by 8 external links, as depicted in the inset of
Fig.~\ref{fig:toric_combined}(d). Suppose we fix the environment as defined by
the gauge spins on the external legs. In the absence of the transverse
field $\widetilde\Gamma$, the star constraint of positive parity is
satisfied, and there are $2^{8-1}=128$ external leg configurations
compatible with the constraint.

In the case of the effective Hamiltonian~(\ref{eq:star}), there are
only two allowed configurations of the free gauge spins satisfying the
star constraints: $|\psi_1\rangle$ and $|\psi_2\rangle$, and they are
related by the plaquette operator: $|\psi_2\rangle = \prod_\square
\sx_i |\psi_1\rangle$. In the presence of a transverse field, the
ground state and the first excited state of this plaquette must be the
symmetric $|\psi_S\rangle$ and antisymmetric $|\psi_A\rangle$
superpositions of $|\psi_1\rangle$ and $|\psi_2\rangle$, respectively;
and the energy splitting between these two states is given by the
energy scale of the effective plaquette operator.

We find for the ``molecule'' Hamiltonian~Eq.~(\ref{eq:toric_code_lattice_exact}) that
the same independence on the external leg configuration holds. In
Fig.~\ref{fig:toric_combined}(d) shows the ground state and the first
excited state energies corresponding to all 128 possible external leg
configurations. In the inset of Fig.~\ref{fig:toric_combined}, we show explicitly three examples of external leg configurations. We find that the energies of both the
ground state and first excited states are exactly the same for all 128
configurations, within machine precision. This is compelling evidence
that the transverse field $\widetilde{\Gamma}$ does not energetically
favor any specific ordering pattern on the lattice. In other words,
the effective plaquette operator generated by the transverse field
$\widetilde{\Gamma}$ does not favor any one spin order in particular,
as it should in a spin liquid.


\subsection*{3D toric code}
\label{sec:3Dtoric}

\begin{figure}[h]
\centering
\includegraphics[width=.45\textwidth]{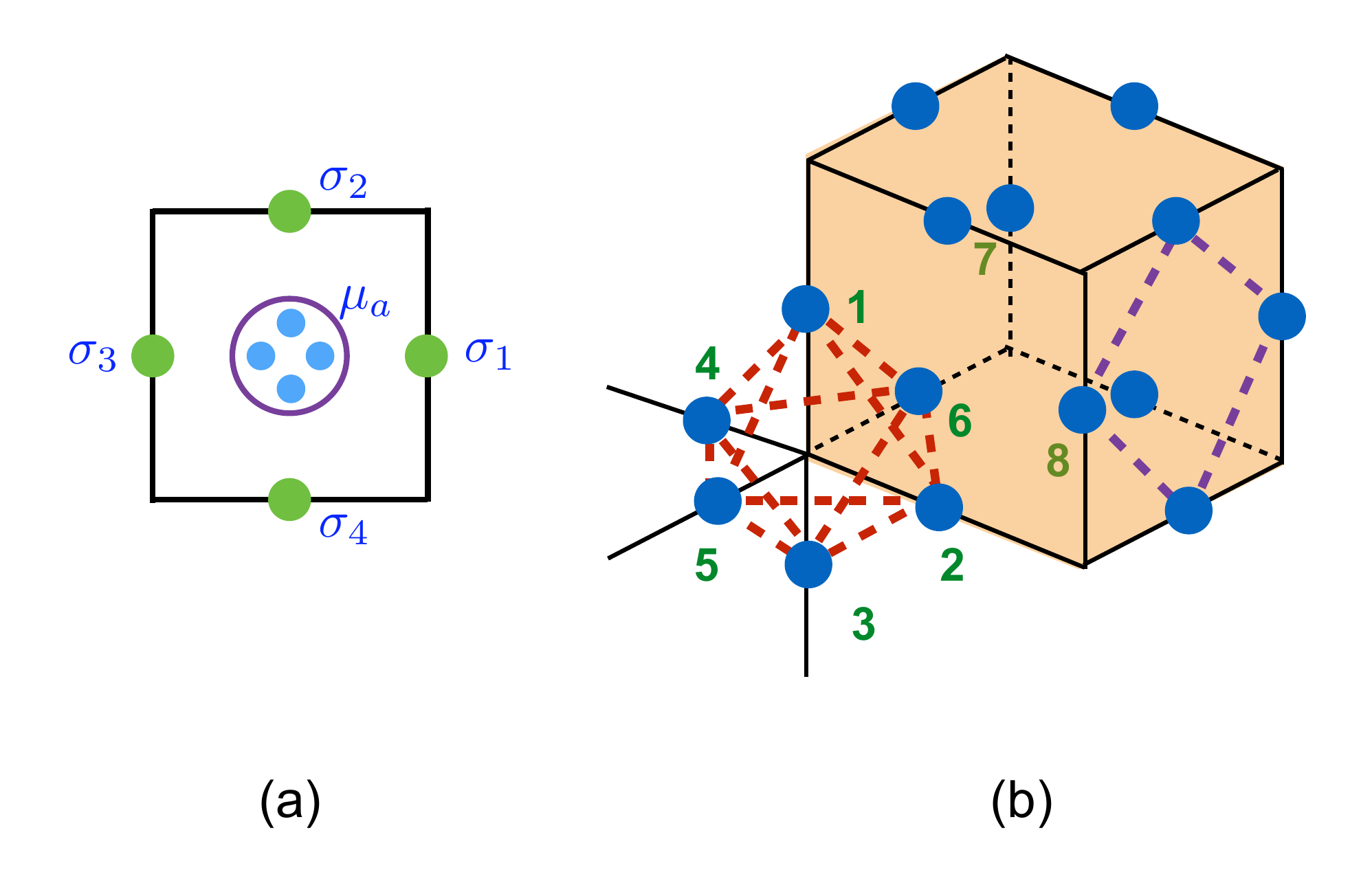}
\caption{Construction of three dimensional toric code model. (a) Matter spins are put at the center of each plaquette. (b) The plaquette operator (depicted in purple line), which is a product of four $\sz$ operators around every face of the cube; and the star operator (depicted in red line), which is a product of six $\sx$ operators on the links emanating from each vertex.}
\label{fig:toric_3D} 
\end{figure}

Following the path that we laid out in two dimensions, we can use Hamiltonian~(\ref{eq:toric_code_lattice_exact}) to generate the toric code in three dimensions, as well. The main difference is that in 3D, we shall put the matter spins at the center of each plaquette (square face) of the cubic lattice, rather than on the vertex, as shown in Fig.~\ref{fig:toric_3D}(a). This will enable us to construct the exact plaquette operator in the $\sz$-basis, first. 
The Hamiltonian for a single plaquette is still given by Eq.~(\ref{eq:star_full}), for which we have shown that the ground state sector takes the form of the product of four $\sz$'s around the plaquette, as in Eq.~(\ref{eq:star}). This yields the exact plaquette term of the 3D toric code. Adding a transverse field $\widetilde{\Gamma}$ on the gauge spins and going to the full cubic lattice, we arrive at the effective Hamiltonian on the full lattice:
\begin{align}
H_{\rm eff}^{\rm 3D-toric}
=
- \lambda \sum_p \;\prod_{i\in p} \sz_i
-\widetilde \Gamma \sum_i \sx_i
\;.
\label{eq:toric_code_lattice_3D}
\end{align}
When $\widetilde{\Gamma}/{\lambda}$ is small, the leading order term acting within the ground state manifold happens at the sixth order, which involves the product of six $\sx$'s on the links emanating from a single vertex. This term is precisely the star operator of the 3D toric code, as shown in Fig.~\ref{fig:toric_3D}(b). Therefore, the same construction can implement the toric code model in both two and three dimensions.

\subsection*{X-cube model}
\label{sec:xcube_lattice}

\begin{figure*}
\centering
\includegraphics[width=\textwidth]{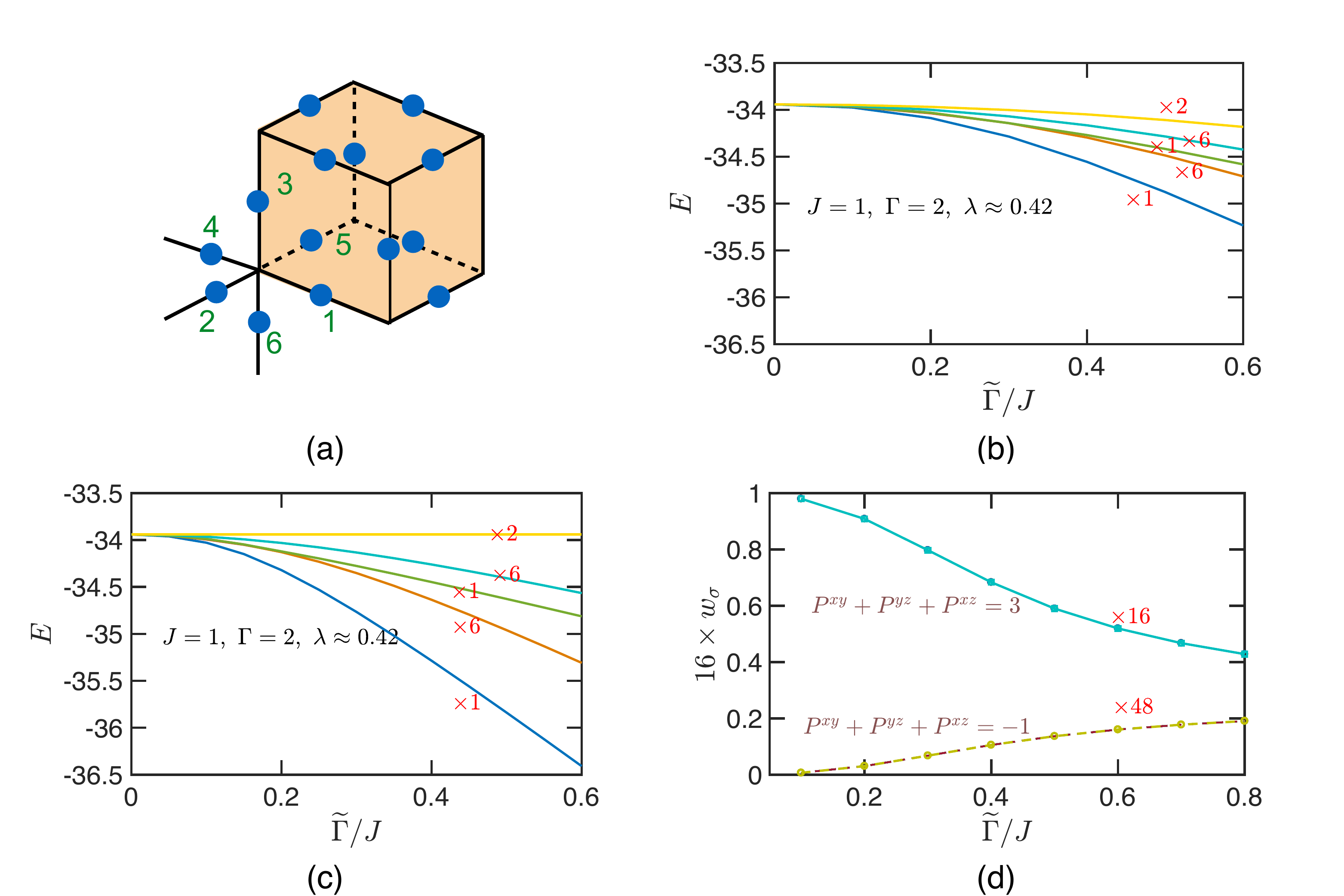}
\caption{(a) The X-cube model exhibiting fractonic topological order. The Hamiltonian contains three star terms associated with three intersecting planes: $B_s^{xy} = \sz_1\sz_2\sz_4\sz_5$, $B_s^{yz}=\sz_2\sz_3\sz_5\sz_6$, and $B_s^{xz}=\sz_1\sz_3\sz_4\sz_6$. The cube operator is the product of $\sx$ around an elementary cube $c$: $A_c=\prod_{n\in \partial c} \sx_n$. (b)\&(c) Eigenenergy spectrum of a single vertex in the X-cube model as a function of the transverse field $\widetilde{\Gamma}$ acting on the gauge spins. Only the levels in the ground state sector where all three star operators have parity $P=+1$ are shown. (b) Energy spectrum of the lowest energy sector of the Hamiltonian with gauge and matter spins. (c) Energy spectrum of the effective Hamiltonian~(\ref{eq:star_xcube}).  Notice that the eigenstates of the Hamiltonian with gauge and matter spins have precisely the same levels of degeneracy as the effective Hamiltonian~(\ref{eq:star_xcube}). (d) Weight of the ground state wavefunction of the full Hamiltonian in the presence of a transverse field $\widetilde{\Gamma}$ on each of the 16 configurations with $P^{xy}+P^{yz}+P^{xz}=3$ (ground states) and the 48 configurations with $P^{xy}+P^{yz}+P^{xz}=-1$ (excited states) of the X-cube star operators, as a function of $\widetilde{\Gamma}$. The curves in each set fall on top of one another, indicating that the ground state in the presence of $\widetilde{\Gamma}$ is an equal amplitude superposition of the configurations with $P^{xy}+P^{yz}+P^{xz}=3$ and those with $P^{xy}+P^{yz}+P^{xz}=-1$, as it should be for the effective Hamiltonian~(\ref{eq:star_xcube}). The degeneracy of each energy level, or the number of curves falling on top of one another is labeled below the curve. The choice of parameters are the same as in Fig.~\ref{fig:toric_combined}.}
\label{fig:xcube_combined} 
\end{figure*}

The X-cube Hamiltonian contains, at each vertex, three star terms
associated with three intersecting planes. Fig.~\ref{fig:xcube_combined}(a)
illustrates the labeling of the 6 spins at the edges of a vertex. The
three stars correspond, each, to a product of 4 spins: $B_s^{xy}
= \sz_1\sz_2\sz_4\sz_5$, $B_s^{yz}=\sz_2\sz_3\sz_5\sz_6$, and
$B_s^{xz}=\sz_1\sz_3\sz_4\sz_6$.

We implement each of the three star operators using the same scheme we
used in the toric code. We group the 6 gauge spins in (overlapping)
sets of 4 spins, forming four-legged stars. The three groups are
$(1245), (2356)$ and $(1346)$, matching the groups in the operators
$B_s^{xy},B_s^{yz}$ and $B_s^{xz}$ above. For each of the different
directions, we need a set of 4 matter spins, thus we require 12 matter
spins per vertex or site of the cubic lattice.  Without the transverse
field $\tilde\Gamma$ on the 6 gauge spins, the three directions are
decoupled, and it thus follows directly from the construction in the
previous section that the low energy effective Hamiltonian takes the form
\begin{align}
H_{\rm eff}^{\rm star}
&=
3\,\gamma - \lambda \;\sz_1\,\sz_2\,\sz_4\,\sz_5 - \lambda \; \sz_2\sz_3\sz_5\sz_6 - \lambda \; \sz_1\sz_3\sz_4\sz_6
\nonumber\\
\nonumber\\
&=
3\,\gamma - \lambda \left(B_s^{xy} + B_s^{yz} + B_s^{xz} \right) 
\;,
\label{eq:star_xcube}
\end{align}
where the coefficients $\gamma$ and $\lambda$ are given in terms of
$J$ and $\Gamma$ by Eq.~(\ref{eqs:gamma-lambda}). The
ground state configuration of each of three star operators $B_s^{xy}, B_s^{yz}$ and $B_s^{xz}$ has
positive parity, $P^{xy} =P^{yz}=P^{xz}=+1$, since $\lambda
>0$. (Notice that there is a constraint that the product
$P^{xy}\,P^{yz}\,P^{xz}=+1$.)

Paralleling the discussion for the 2D $\mathbb{Z}_2$ gauge theory,
there is a regime where $|\Gamma|\to \infty$ while keeping $\lambda$
fixed, which opens an infinite gap to the excited sectors. The splitting $2|\lambda|$ between the two
parity states within the lowest energy sector remains finite. To
access this regime we would fix $\lambda$ and tune
$J=\left|\lambda\,\Gamma^3/12\right|^{1/4}$ just as in the 2D case.

When $\widetilde\Gamma=0$, the spectrum of the gauge-matter
Hamiltonian (with six gauge spins and three sets of four matter spins)
depends only on the {\it sum} of the parities
$P^{xy}+P^{yz}+P^{xz}$. This is a manifestation of the combinatorial
gauge symmetry we identified in the previous section, and this symmetry is
again essential to suppressing any interaction that favors any type of
order.

So far we have obtained the exact star operators of a single vertex in the X-cube model. Now consider the full cubic lattice in three dimensions where the gauge spins reside on the links and the matter spins reside on the vertices. We apply a transverse field $\widetilde{\Gamma}$ to the gauge spins.

First, consider the limit $|\Gamma|\rightarrow \infty$, $J=|\lambda \Gamma^3/12|^{1/4}$ with $\lambda$ fixed (i.e., projecting down to the ground state sector). In this limit, the Hamiltonian on the entire lattice becomes:
\begin{equation}
H_{\rm eff}^{\rm X-cube} = -\lambda \sum_s \left(B_s^{xy} + B_s^{yz} + B_s^{xz} \right) -\widetilde{\Gamma} \sum_i \sx_i.
\label{eq:xcube}
\end{equation}
In the limit where $\widetilde{\Gamma}/\lambda$ is small, the lowest order term in perturbation theory that acts within the ground state subspace is the 12-body interaction around each cube $c$: $A_c = \prod_{n\in \partial c} \sx_n$, which is the cube operator of the X-cube model. Similar to Hamiltonian~(\ref{eq:toric_code_lattice}), we expect that there is a range of small $\widetilde{\Gamma}/\lambda$ where the system is in the phase with fractonic topological order. We have thus implemented the full X-cube model on the lattice, using only two-body Ising couplings and a transverse field.

Similar to the implementation of toric code in 2D, we shall now provide numerical evidence indicating that the full Hamiltonian with gauge and matter spins preserves all essential symmetries of the exact star operators in the X-cube model for a wide range of parameters $(\widetilde\Gamma, \Gamma, J)$. Since there are now three star operators in the X-cube model satisfying the constraint $P^{xy}P^{yz}P^{xz}=+1$, the ground states must have parity +1 for all three star operators, or equivalently, $P^{xy}+P^{yz}+P^{xz}=3$; the excited states are created by flipping the parities of two out of the three stars, i.e. $P^{xy}+P^{yz}+P^{xz}=-1$. In Fig.~\ref{fig:xcube_combined}(b)\&(c), we plot the eigenenergy spectrum of the ground state sector where $P^{xy}+P^{yz}+P^{xz}=3$, in the presence of a transverse field $\widetilde{\Gamma}$ on the gauge spins. Once again, the levels of the full Hamiltonian [Fig.~\ref{fig:xcube_combined}(b)] have a one-to-one correspondence to the levels of the exact star operators [Fig.~\ref{fig:xcube_combined}(c)] for nonzero $\widetilde{\Gamma}$, with precisely the same degeneracies. While the levels of the excited states are quite complicated to count directly, transitions to the excited states can be captured again by looking at the weight of the ground state wavefunction on every classical configuration, including excited states with $P^{xy}+P^{yz}+P^{xz}=-1$, just like in Fig.~\ref{fig:toric_combined} for the 2D toric code. This is shown in Fig.~\ref{fig:xcube_combined}(d). It is clear that the ground state wavefunction has equal weights on each of the 16 configurations with $P^{xy}+P^{yz}+P^{xz}=3$ (ground states) and the 48 configurations with $P^{xy}+P^{yz}+P^{xz}=-1$ (excited states), as should be the case for the exact star operators of the X-cube model. While we are unable to perform exact diagonalizations of a single cube surrounded by fixed external legs, as we did in Fig.~\ref{fig:toric_combined}(d) for the toric code in 2D, our numerical results suggest that we indeed replicate a single vertex in the X-cube model even when quantum dynamics is introduced.

\end{document}